\def\draftversion{false} 
\def\showall{true}  

\RequirePackage{ifthen}
\ifthenelse{\equal{\draftversion}{true}}{
  \documentclass[aps,prb,10pt,galley,amsmath,amssymb,
                 longbibliography,superscriptaddress,
                 nofootinbib,notitlepage,
                 showpacs,preprintnumbers]{revtex4-1}
\usepackage{showlabels}
}{
  \documentclass[aps,prb,10pt,twocolumn,amsmath,amssymb,
    longbibliography,superscriptaddress,
    showpacs,preprintnumbers]{revtex4-1}
}

\usepackage{MSGstyle}
\usepackage{my_comment}

\begin{document}

\title{Strain-induced collapse of Landau Levels in real Weyl semimetals}

\author{Yang-Jun Lee}
\email{dldidwns1123@snu.ac.kr}
\affiliation{Center for Correlated Electron Systems, Institute for Basic Science, Seoul 08826, Korea}
\affiliation{Department of Physics and Astronomy, Seoul National University, Seoul 08826, Korea}
\affiliation{Center for Theoretical Physics, Seoul National University, Seoul 08826, Korea}

\author{Cheol-Hwan Park}
\email{cheolhwan@snu.ac.kr}
\affiliation{Center for Correlated Electron Systems, Institute for Basic Science, Seoul 08826, Korea}
\affiliation{Department of Physics and Astronomy, Seoul National University, Seoul 08826, Korea}
\affiliation{Center for Theoretical Physics, Seoul National University, Seoul 08826, Korea}
\affiliation{Donostia International Physics Center, 20018 San Sebastián, Spain}
\affiliation{Centro de Física de Materiales, Universidad del País Vasco, EHU, 20018 San Sebastián, Spain}

\author{Mar\'ia A. H. Vozmediano}
\email{vozmediano@icmm.csic.es}
\affiliation{Instituto de Ciencia de Materiales de Madrid, CSIC, Cantoblanco, Madrid 28049, Spain}

\date{\today}

\begin{abstract}
The collapse of Landau levels under an electric field perpendicular to the magnetic field is one of  the distinctive features of Dirac materials. So is the coupling of lattice deformations to the electronic degrees of freedom in the form of gauge fields which allows the formation of pseudo-Landau levels from strain. We analyze the collapse of Landau levels induced by  strain on realistic Weyl semimetals hosting anisotropic, tilted Weyl cones in momentum space. We perform first-principles calculations, to establish the conditions on the external strain for the collapse of Landau levels in TaAs which can be experimentally accessed.
\end{abstract}

\maketitle

\section{INTRODUCTION}
Dirac materials, such as graphene~\cite{neto2009electronic} and Weyl semimetals (WSMs)~\cite{BF17,AMV18}, are characterized by having the fermi level near a pair of band crossings in momentum space. The dispersion relation of the low energy electronic excitations around the Fermi level is linear and the quasiparticles are described by the relativistic, massless Dirac equation. This fact is at the origin of the recent fascination for the Dirac materials, especially after the synthesis of the three dimensional Dirac and Weyl semimetals~\cite{lv2015observation,lv2015experimental,JXH16,BF17,AMV18}.  Phenomena described in the formalism of quantum field theory as the chiral or gravitational anomalies are being observed in the material realization of Weyl physics~\cite{CFetal21}.


The structure of the Landau levels (LLs) under a magnetic field is one of the distinctive characteristics of the Dirac spectrum. The dependence of the cyclotron frequency on the magnetic field and the spacing of the LLs are different than these of the standard non-relativistic electron systems, and, more importantly, there is a zeroth LL with linear dispersion  band~\cite{TCG16}. In three spacial dimensions, the zeroth  LL plays an important role in the realization of the chiral anomaly, the non-conservation of chiral charge when \textit{parallel} electric and magnetic fields are applied~\cite{NN83,burkov2015chiral,JXH16}.


An important relativistic effect of the Dirac system occurs when an electric field is applied \textit{perpendicularly} to the magnetic field:   It has been shown that, at a critical value of the electric field, the LLs collapse and the spectrum becomes continuus again. The collapse of LLs  was first predicted to occur in graphene~\cite{lukose2007novel,Peres_2007, nimyi2022landau} and the possible  experimental confirmations were analyzed in Refs.~\cite{VD2009,KL2011}. This phenomena was extended to Weyl semimetals in Ref.~\cite{arjona2017collapse} and to Kane fermions  in Ref.~\cite{krishtopenko2021relativistic}

Another interesting phenomena arising in Dirac matter is the fact that lattice deformations couple to the electronic density in the form of pseudo-electromagnetic gauge potentials \cite{cortijo2015elastic,CKLV16,Strain20,Strain21}.  Pseudomagnetic fields induce LLs~\cite{StrainLL19} and the additional inclusion of perpendicular pseudo-electric field can lead to the collapse of these LLs~\cite{arjona2017collapse}. The possible
  experimental observation of this phenomenon would be a direct confirmation of the reality of pseudo-gauge fields with important potential applications~\cite{Strain20,Strain21}.
  
 The collapse of LLs in WSMs induced by elastic pseudo-electromagnetic fields was analyzed in Ref.~\cite{arjona2017collapse} for the ideal case of a WSM having a pair of  isotropic, non-tilted Weyl nodes. However, real    materials  such as TaAs have {tilted} and anisotropic Weyl cones in the bandstructure~\cite{lv2015observation}, and  more complicated models to study the collapse of LLs were proposed in Refs.~\cite{jafari,alisultanov}. Still, these models are based on the effective low-energy continuum description with tunable parameters, not allowing a comparison with possible experiments. Modelling the Hamiltonian from the results of first-principles calculations on real WSMs is an important step towards the experimental confirmation of this novel phenomenon.
 
In this work, we will first review the criterion for the collapse of LLs for a tilted and isotropic Weyl cone in Sec.~\ref{sec:LLcollapse}. In Sec.~\ref{sec:strain_collapse_LLs}, we will introduce the strain-induced pseudo-electromagnetic fields and review the condition for the collapse of LLs at a tilted, isotropic Weyl cone similarly as in Refs.~\cite{jafari,alisultanov}. We extend the analysis including both the tilt of the Weyl cone and anisotropy in the velocity in Sec.~\ref{sec:anisotropy}.  Based on the developed theory and first-principles calculations, we will present the criterion for the strain-induced collapse of LLs in TaAs in Sec.~\ref{sec:TaAs}. We summarize our work and discuss open problems in Sec.~\ref{sec:conclusions}.

\section{COLLAPSE OF LANDAU LEVELS IN a WEYL SEMIMETAL} \label{sec:LLcollapse}
For completeness and to fix the notation,  we will describe in this section the collapse of LLs in WSMs with and without tilted cones discussed previously in Refs.~\cite{arjona2017collapse,alisultanov}. 
The low-energy Hamiltonian of a WSM around a Weyl node with chirality $\chi=\pm 1$ is
\begin{align} \label{eq:hamiltonian}
\begin{split}
  \mathcal{H}=\chi\, v_{\rm F}\, \sigma\cdot \textbf{p}\,,
 \end{split}
\end{align}
where $v_{\rm F}$ is the Fermi velocity, $p_i$ the momentum operator ($i=1,2,3$), and $\sigma^i$'s are the Pauli matrices.

If a magnetic field $\textbf{B}=B_z\, \hat{{z}}$ associated to the vector potential $\textbf{A}=(-B_z y, 0, 0)$  in  the Landau gauge is applied, the spectrum of the Hamiltonian organizes into LLs~\cite{rabi1928freie}:
\begin{align} \label{eq:lls}
\begin{split}
  \varepsilon^{\chi}_{n}=\pm v_{\rm F} \sqrt{2 l_{B}^{-2} n +  k_{z}^2}\,\,\,\,(n=1,\,2,\,3,\,...)
 \end{split}
\end{align}
where we set $\hbar=1$, $l_{\rm{B}} = \sqrt{v_{\rm F}/eB_z}$ is the magnetic length, $-e$ is the electron charge, and $k_z$ is the momentum in the direction of the magnetic field. Note that LLs of WSM is are proportional to $\sqrt{n}$, whereas the LLs of ordinary materials are proportional to $n$. The chiral zeroth LL, a characteristic of the Dirac system, is described by
\begin{align} \label{eq:lls0th}
\begin{split}
  \varepsilon^{\chi}_{0}=\chi\, v_{\rm F} k_{z}\,.
 \end{split}
\end{align}

An electric field $\textbf{E}=E_y \hat{y}$ is now applied perpendicularly to the magnetic field $\textbf{B}=B_z \hat{z}$ . A very elegant solution to this problem uses the fact that  the low-energy states around the Weyl node are Lorentz invariant with the speed of light $c$ being replaced with the Fermi velocity $v_{\rm{F}}$\cite{lukose2007novel}. Therefore, a boost transformation of the fields with velocity ${\bf v}=v_x\hat{x}$ leads to~\cite{jackson1999classical}
\begin{align} \label{eq:lboostE}
\begin{split}
  E'_y &= \gamma (E_y - v_x B_z)
 \end{split}\\\label{lboostB}
  B'_z &= \gamma \left(B_z - \frac{v_x}{v^2_{\rm{F}}} E_y\right)\,,
\end{align}
where $\gamma\equiv1/\sqrt{1-\beta^2}$ and $\beta\equiv v_x/v_{\rm F}$.
The electric field in the moving reference frame is  zero if $v_x=v_0$ where
\begin{equation} \label{eq:v_0}
    v_0 \equiv E_y / B_z\,,
\end{equation}
or, equivalently, if $\beta = E_y / v_{\rm{F}}B_z$.

We can then write the  LLs in the moving frame with $E'_y=0$, $\varepsilon'^{\chi}_{n}$, and  find LLs in the lab frame by  inverting the Lorentz transformation. The final result is
\begin{align} \label{eq:lls2}
\begin{split}
  \varepsilon^{\chi}_{n}&=\gamma(1-\beta^2)\varepsilon'^{\chi}_{n}+v_{\rm{F}} \beta k_x\\
  &=\pm \sqrt{2v_{\rm{F}}^{2} l_{B}^{-2} n (1-\beta^2)^{3/2}+ v_{\rm{F}}^{2} k_{z}^2(1-\beta^2)}+v_{\rm{F}} \beta k_x\,.
 \end{split}
\end{align}
From Eq.~\eqref{eq:lls2} we can see that the LLs collapse at the critical value $\beta \geq 1$ 
or $E_y \geq  v_{\rm{F}} B_z$. It is important to note that the LL collapse does not occur in a non-relativistic electron gas, making it  a distinctive property of Dirac materials, such as WSMs or graphene.

Real WSMs have tilted Weyl cones described by the following Hamiltonian:
\begin{align} \label{eq:tilt}
\begin{split}
 \mathcal{H}=\textbf{w} \cdot \textbf{p}+\chi\, v_{\rm{F}}\, \sigma\cdot \textbf{p}\,,
 \end{split}
\end{align}
where $\textbf{w}$ is the tilt velocity.
This Hamiltonian does not have an analogy in special relativity, so  alternative methods are needed to obtain the criterion for the collapse of LLs. This problem has been addressed with    general relativity techniques~\cite{jafari} and with algebraic methods~\cite{alisultanov}.  The modified LLs for the tilted cone  given in Ref.~\cite{alisultanov} are
\begin{align} \label{eq:lls_tilt}
\begin{split}
  \varepsilon^{\chi}_{n}=\pm \sqrt{2 \gamma^{-3}_{\chi} v_{\rm{F}}^{2} l_{B}^{-2} n + \gamma^{-2}_\chi v_{\rm{F}}^{2} k_{z}^2}+w_z k_z + v_0 k_x
 \end{split}
\end{align}
for a positive integer $n$ where $$\gamma_\chi = 1/\sqrt{1-[(\chi v_0- w_x)^2 +w_y ^2]/v^2 _{\rm{F}}}.$$ The modified criterion for the collapse of LLs is
\begin{align} \label{eq:crit_tilt}
\begin{split}
  (\chi v_0 - w_x)^2 +w_y ^2 \geq v_{\rm{F}} ^2 ,
 \end{split}
\end{align}
where, as in the previous case, $v_0=E_y/B_z$.

In real WSMs, in addition to the tilt of the Weyl cones, the Fermi velocity is also anisotropic. We will extend the formalism to the realistic, general case later.

\section{STRAIN-INDUCED COLLAPSE OF LANDAU LEVELS IN a WEYL SEMIMETAL} \label{sec:strain_collapse_LLs}

The fact that
elastic deformations of the lattice couple to the electronic Hamiltonian of Dirac matter as elastic gauge fields was first recognized in graphene where it gave rise to a new line of research called  straintronics~\cite{Amorim16, suzuura2002phonons}. More recently, the attention has moved to the strain-induced gauge fields in WSMs~\cite{cortijo2015elastic,CKLV16} and the physical consequences of the pseudo-electromagnetic fields ~\cite{Strain20,Strain21}. 

When materials are strained, the hoping parameters between atomic orbitals and on-site energies are both changed. In linear elasticity theory~\cite{Landau} the main role is played by the strain tensor defined as $u_{ij}=1/2(\partial u_i/\partial x^j+\partial u_j/\partial x^i)$  where $u_i$ is the deformation vector.  For small elastic deformations  we can assume that the change in the Hamiltonian due to the deformation depends linearly  on $u_{ij}$. If  the Weyl nodes are slightly shifted in the Brillouin zone, this Weyl node shift can be interpreted as a \textit{pseudo-magnetic gauge field} due to the strain \cite{cortijo2015elastic}. 

Here, to adapt to the {\it ab initio} calculations to be introduced later, we will propose a formulation that differs from the previous ones described in Ref.~\cite{cortijo2015elastic}. We call the ratio between the Weyl node shift and strain tensor \textit{the Weyl node shift per unit strain}. 
The deformation of the crystal lattice can also shift the energy position of each band, 
and we will call the ratio between the shift in energy with respect to the Fermi level and strain \textit{the energy shift per unit strain}.
\begin{figure}
    \includegraphics[width=0.50\textwidth]{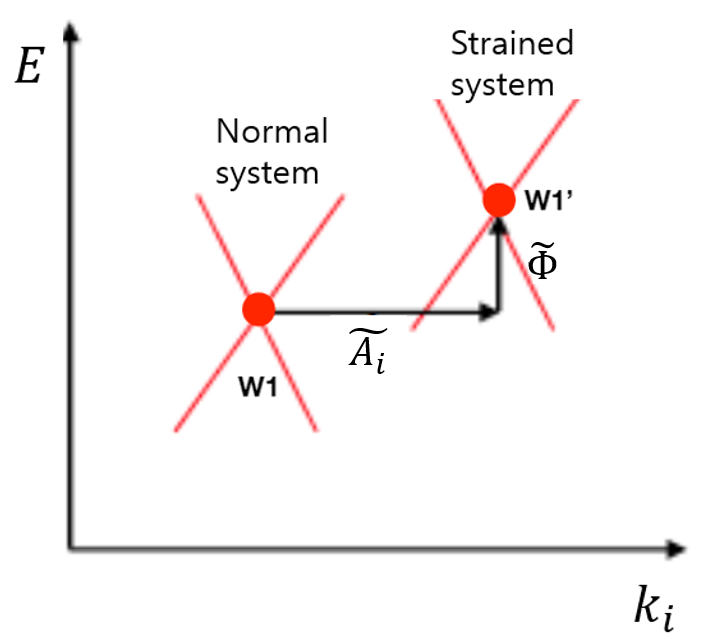}
    \caption{Shift of the position of a single Weyl node $W_1$ to $W_1'$ under strain defining the pseudo-gauge fields in eqs \eqref{a_i} and~\eqref{phi}.}
    \label{fig:strain}
\end{figure}
We can incorporate these two effects into the Hamiltonian as follows [Fig.~\ref{fig:strain}]:
\begin{align} \label{eq:hamiltonian2}
\begin{split}
  \mathcal{H}={\chi}\,&v_{\rm{F}}\, \sigma\cdot (\textbf{p}+\tilde{\textbf{A}})+\tilde{\Phi}
   \end{split}
\end{align}
with
\begin{align} \label{a_i}
\begin{split}
  \tilde{A_i}=& \sum_{jk} b_{ijk} u_{jk}
   \end{split}\\\label{phi}
   \tilde{\Phi} =& \sum_{ij} g_{ij} u_{ij}.
\end{align}
Here, we used symbols with tilde, $\tilde{A_i}$ and $\tilde{\Phi}$, to denote that they are not the actual vector and gauge potentials, respectively, but are strain-induced {\it pseudo} quantities.

As we see, the Weyl node shift per unit strain $b_{ijk}$ is a rank three tensor that defines the magnitude and direction of the node shift $\tilde{\bf A}$ under strain.  It encodes all the  material's details (anisotropy, elastic parameters and alike). Similarly, in the deformation potential associated to the  displacement of the position in energy of the Weyl node, we absorb the material parameters in  the energy shift per unit strain $g_{ij}$. These tensors will be obtained from first-principles calculations. 
 The present formulation,  based on the displacement of a single Weyl node under strain is, in principle,  more general than the  standard formulation in the continuum  limit \cite{cortijo2015elastic} based on the vector separating the two Weyl nodes. 

The strain-induced gauge field and the energy shift per unit strain couple to the electronic degrees of freedom as the electromagnetic vector and   scalar potentials, respectively. Hence, we can introduce the strain-induced \textit{pseudo-electromagnetic} fields as
\begin{align} \label{bfield}
\begin{split}
  \tilde{\textbf{B}} &= \nabla \cross \tilde{\textbf{A}}
   \end{split}\\\label{efield}
   \tilde{\textbf{E}} &= -\nabla \tilde{\Phi}\,.
\end{align}
Note that strain-induced pseudo-gauge fields differ from real electromagnetic gauge fields in that they may be different for different Weyl nodes. For example, for a time-reversal-symmetric WSM such as TaAs, the pseudo-gauge fields near the time-reversal paired nodes are of opposite signs.

Next, we will describe the condition under which  strain will induce the collapse of LLs. 
Since the elastic fields \eqref{bfield} involve derivatives of the strain tensor, only inhomogeneous strain configurations will give rise to non trivial physical effects. Consider the simple case where the only non-zero  component of the strain tensor is 
\begin{equation} \label{eq:u_xx}
u_{xx}=\textbf{a}\cdot\textbf{x}\,.
\end{equation}
Here, $\textbf{a}$ is the constant {\it strain gradient vector} at position $\textbf{x}$. This vector, together with the strain of Weyl node shift  will determine the collapse of the LLs. 
Using the strain gradient vector, we can write the Weyl node shift and the energy shift per unit strain as
\begin{align} \label{a2}
\begin{split}
  \tilde{\textbf{A}} &=\textbf{b}\, u_{xx} = \textbf{b}\, (\textbf{a}\cdot \textbf{x})
   \end{split}\\\label{phi2}
   \tilde{\Phi} &= g\, (\textbf{a}\cdot \textbf{x})\,,
\end{align}
where,  in this simple case, $b_i$ in Eq.~\eqref{a2} is $b_{ixx}$ in Eq.~\eqref{a_i}, $g_{xx}$ in Eq.~\eqref{phi} is $g$ in Eq.~\eqref{phi2}, and all the other $b_{ijk}$ and $g_{ij}$ components are zero.

We will also consider later  the case where the only non-zero strain component is given by
\begin{equation} \label{eq:u_zz}
u_{zz}=\textbf{a}\cdot\textbf{x}\,.
\end{equation}
The quantities ${\bf b}$, $g$, $\tilde{\bf A}$ and $\tilde{\Phi}$ are similarly defined as in the previous case.  We will use more general strain configurations in sec. \ref{sec:TaAs}.

From Eqs.~\eqref{a2} and~\eqref{phi2}, the pseudo electromagnetic fields read
\begin{align} \label{bfield2}
\begin{split}
  \tilde{\textbf{B}} &= \nabla\cross\tilde{\textbf{A}}= \textbf{a} \cross \textbf{b}
   \end{split}\\\label{efield2}
   \tilde{\textbf{E}} &= -\nabla \tilde{\Phi}= -g\,\, \textbf{a}\,.
\end{align}

Consequently, the criterion for the LL collapse [Eq.~\eqref{eq:crit_tilt}] for this particular example  becomes
\begin{align} \label{crit2}
\begin{split}
   \frac{|\textbf{b}|^2}{g^2} \left[ ( v_{\rm{F}}^2 - w_x^2 - w_y^2 ) \,\textrm{sin}^2(\theta_{\textbf{a},\textbf{b}}) - 2\chi\,\frac{g w_x}{|\textbf{b}|}\, \textrm{sin}(\theta_{\textbf{a},\textbf{b}}) \right]\leq1
   \end{split}
\end{align}
where $\theta_{\textbf{a,b}}$ is the angle between the Weyl node shift per unit strain $\textbf{b}$ and the strain gradient vector $\textbf{a}$.

Here, we assume the $x$ and $y$ axes to be along $\tilde{\textbf{E}}\cross\tilde{\textbf{B}}=-g\,\, \textbf{a}\cross(\textbf{a} \cross \textbf{b})$ and $\tilde{\textbf{E}}=-g\,\textbf{a}$, respectively. The $( v_{\rm{F}}^2 - w_x^2 - w_y^2 )$ term could be negative in the case of type II Weyl semimetals, but it is always positive in the case of type I Weyl semimetals\cite{AMV18}. Note that the criterion for the LL collapse  does not depend on the magnitude of the strain gradient vector $|{\bf a}|$. This is similar to the result found in refs. \cite{castro2017raise,arjona2017collapse} where the conditions for the LL collapse turned into conditions on the elastic coupling constants of the material.
Notice also that, if the cones are not tilted, the condition for the collapse is the same for the two Weyl nodes of opposite chirality.

\section{Generalization to anisotropic Weyl cones} \label{sec:anisotropy}

The Weyl cone in the bandstructure of a real WSM has both tilt and anisotropy, resulting in an effective Hamiltonian of the following form:
\begin{align} \label{eq:anisotropy0}
\begin{split}
\mathcal{H}=w^i p_i+\sigma^i v_{ij}p_j
 \end{split}
\end{align}
or, using matrix and vector notations,
\begin{align} \label{eq:anisotropy}
\begin{split}
 \mathcal{H}=\textbf{w} \cdot \textbf{p}+ \sigma \cdot \textbf{v} \, \cdot \textbf{p},
 \end{split}
\end{align}
where $v_{ij}$ is the matrix element for anisotropic Fermi velocities.
One can directly obtain the velocity operator projected onto the double-degenerate space
\begin{equation} \label{eq:vel_op}
\frac{d\mathcal{H}}{d\textbf{p}}=\textbf{w}+\sigma \cdot \textbf{v}
\end{equation}
by using a Wannier-function-based method (see Supp. Info. of Ref.~\cite{kim2017breakdown}).
Instead, we use the energy dispersion to extract the necessary information on $\textbf{w}$ and $\textbf{v}$.
Using the polar decomposition theorem, we can uniquely decompose the real matrix $\textbf{v}$ as
\begin{equation}
\textbf{v}=\textbf{O}\,\textbf{U}
\label{eq:v_OU}
\end{equation}
where $\textbf{O}$ is a real, orthogonal $3\times3$ matrix and $\textbf{U}$ a real symmetric, positive semidefinite $3\times3$ matrix. We can further decompose $\textbf{O}$ as $\textbf{O}=\chi\,\textbf{R}$ where $\chi=\pm1$ is the chirality and $\textbf{R}$ is a proper rotation matrix whose determinant is $+1$. $\textbf{U}$ can be further diagonalized as
\begin{equation}
\textbf{U}=\tilde{\textbf{R}}\textbf{D}\tilde{\textbf{R}}^{\rm T}
\label{eq:U}
\end{equation}
where $\tilde{\textbf{R}}$ is another proper rotation matrix and $\textbf{D}$ is a diagonal velocity matrix whose diagonal components are non-negative and are denoted as $v_x'$, $v_y'$, and $v_z'$:
\begin{equation}
    \textbf{D}=
    \begin{pmatrix}
v_x' & 0 & 0\\
0 & v_y' & 0\\
0 & 0 & v_z'
\end{pmatrix}\,.
\label{eq:D}
\end{equation}

Therefore,
\begin{equation} \label{eq:v}
    \textbf{v}=\chi\,\textbf{R}\,\textbf{U}=\chi\,\textbf{R}\,\tilde{\textbf{R}}\textbf{D}\tilde{\textbf{R}}^{\rm T}\,.
\end{equation}
Using Eq.~\eqref{eq:v}, we can rewrite Eq.~\eqref{eq:anisotropy} as
\begin{equation} \label{eq:anisotropy2}
 \mathcal{H}=\textbf{w} \cdot \textbf{p}+ \chi\,\sigma \cdot \textbf{R}\textbf{U} \textbf{p}\,,
\end{equation}
whose energy eigenvalue is given by
\begin{equation} \label{eq:eigenvalue}
\begin{split}
    \varepsilon_{s\textbf{p}} &= \textbf{w} \cdot \textbf{p}
    +s\sqrt{\textbf{RUp}\cdot\textbf{RUp}}\\
    &=\textbf{w} \cdot \textbf{p}
    +s\sqrt{\textbf{p}^{\rm T} (\textbf{U}^{\rm T}\textbf{U})\textbf{p}}\\
    &=\textbf{w} \cdot \textbf{p}
    +s\sqrt{\textbf{p}^{\rm T} \textbf{U}^2\textbf{p}}\,,
\end{split}
\end{equation}
where $s=\pm1$ is the band index.
Note that the matrix $\textbf{U}^2$ is,  as $\textbf{U}$, symmetric and positive semidefinite. From the energy bandstructure near the Weyl node, we can uniquely determine $\textbf{U}^2$, or, equivalently, $\textbf{U}$ and $\textbf{w}$. Then, by diagonalizing $\textbf{U}$, we obtain both $\tilde{\textbf{R}}$ and $\textbf{D}$ [Eq.~\eqref{eq:U}].

We have to note however that from this method we cannot determine the chirality $\chi$ or the rotation matrix $\textbf{R}$ in Eq.~\eqref{eq:v}. They can also be uniquely determined if we use the Wannier-function-based method developed in Ref.~\cite{kim2017breakdown} (see Supp. Info. therein).

Now, let us take the strain-induced fields into account using Eqs.~\eqref{a2} and~\eqref{phi2}. The Hamiltonian becomes
\begin{equation}  \label{eq:H_complex}
    \begin{split}
        \mathcal{H} &= \textbf{w} \cdot \textbf{p}+\sigma \cdot \textbf{v} \cdot \left[\textbf{p}+\textbf{b}\,( \textbf{a}\cdot \textbf{x})\right]+ g\, (\textbf{a}\cdot \textbf{x})\\
        &=\textbf{w} \cdot \textbf{p}+g\, (\textbf{a}\cdot \textbf{x})\\
        &+\sigma \cdot \left(\chi\, \textbf{R}\,\tilde{\textbf{R}}\textbf{D}\tilde{\textbf{R}}^{\rm T}\right) \left[\textbf{p}+\textbf{b}\,( \textbf{a}\cdot \textbf{x})\right]\\
        &=\textbf{w}' \cdot \textbf{p}'+g\, (\textbf{a}'\cdot \textbf{x}')\\
        &+\chi\, \left(\tilde{\textbf{R}}^{\rm T}\textbf{R}^{\rm T}\sigma\right) \cdot \left[\textbf{p}'+\textbf{b}'\,( \textbf{a}'\cdot \textbf{x}')\right]\,,
    \end{split}
\end{equation}
where we have defined
\begin{equation} \label{eq:primed_vectors}
    \begin{split}
        \textbf{p}'&=\textbf{D}\tilde{\textbf{R}}^{\rm T}\textbf{p}\,,\\
        \textbf{x}'&=\textbf{D}^{-1}\tilde{\textbf{R}}^{\rm T}\textbf{x}\,,\\
        \textbf{w}'&=\textbf{D}^{-1}\tilde{\textbf{R}}^{\rm T}\textbf{w}\,,\\
        \textbf{a}'&=\textbf{D}\tilde{\textbf{R}}^{\rm T}\textbf{a}\,,\\
        \textbf{b}'&=\textbf{D}\tilde{\textbf{R}}^{\rm T}\textbf{b}\,.
    \end{split}
\end{equation}
Note that ${\bf x}'$ and ${\bf p}'$ satisfy the canonical commutation relation:
\begin{equation}
    [x'_i,\, p'_j] = i\hbar\,\delta_{i,j}\,.
\end{equation}

In finding the condition for the LL collapse, we are entitled to perform a proper spin rotation to $\mathcal{H}$ in Eq.~\eqref{eq:H_complex} and finally obtain
\begin{align}  \label{eq:H_complex2}
        \mathcal{H}' &=\textbf{w}' \cdot \textbf{p}'+\chi\, \sigma\cdot \left[\textbf{p}'+\textbf{b}'\,( \textbf{a}'\cdot \textbf{x}')\right]+g\, (\textbf{a}'\cdot \textbf{x}')\,.
\end{align}
Note that now we don't have to worry about the $\textbf{R}$ matrix [Eqs.~\eqref{eq:v} and~\eqref{eq:anisotropy2}] which cannot be obtained from the electronic bandstructure alone. Instead, we should use our knowledge on the chirality $\chi$ of each Weyl node.

Since Eq.~\eqref{eq:H_complex2} is the Hamiltonian for an isotropic Weyl cone with a tilt under strain, we can now use the criterion in Eq.~\eqref{crit2} except that $v_{\rm F}$ is set to $1$ and $\textbf{w}$, $\textbf{a}$, and $\textbf{b}$ are replaced with $\textbf{w}'$, $\textbf{a}'$, and $\textbf{b}'$, respectively.
Finally, the criterion for LL collapse [Eq.~\eqref{crit2}] then reduces to
\begin{align} \label{eq:crit_principal}
\begin{split}
\frac{|\textbf{b}'|^2}{g^2} \left[ ( 1 - {w'_x}^2 - {w'_y}^2 ) \textrm{sin}^2(\theta_{\textbf{a}',\textbf{b}'}) - 2\chi\frac{g\, w'_x}{|\textbf{b}'|} \textrm{sin}(\theta_{\textbf{a}',\textbf{b}'}) \right]\leq1\,.
   \end{split}
\end{align}
\section{Application to $\textbf{TaAs}$} \label{sec:TaAs}

\begin{figure}
    \includegraphics[width=0.45\textwidth]{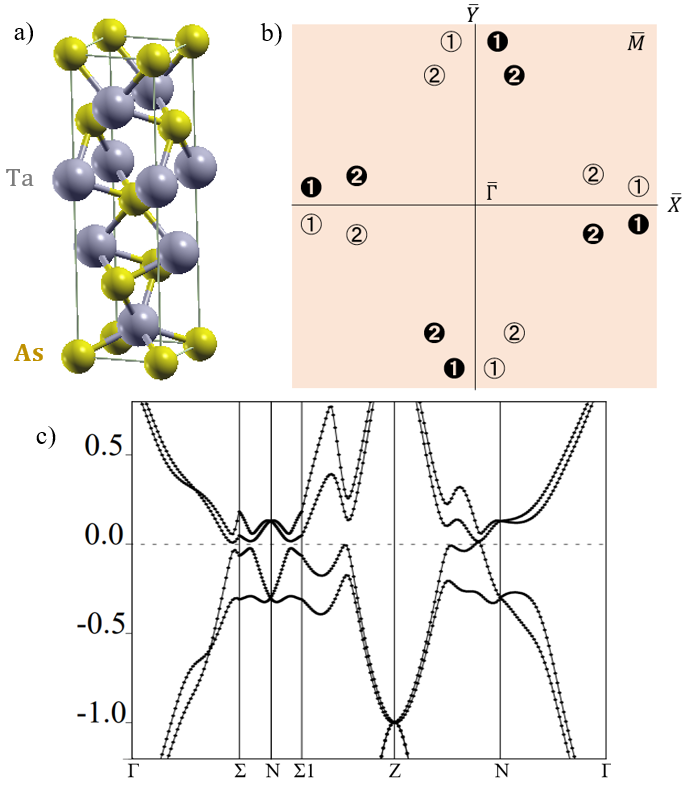}
    
    \caption{(a) Crystal structure of TaAs (b) A Schematic positions of the Weyl nodes of TaAs projected on the (001) surface in the Brillouin zone. There are two types of Weyl nodes: W1 ($k_z=2\pi / c$) and W2 ($k_z \neq 2\pi / c$). (c) The band structure near the Fermi energy}
    \label{fig:structure}
\end{figure}

\begin{figure}
    \includegraphics[width=0.45\textwidth]{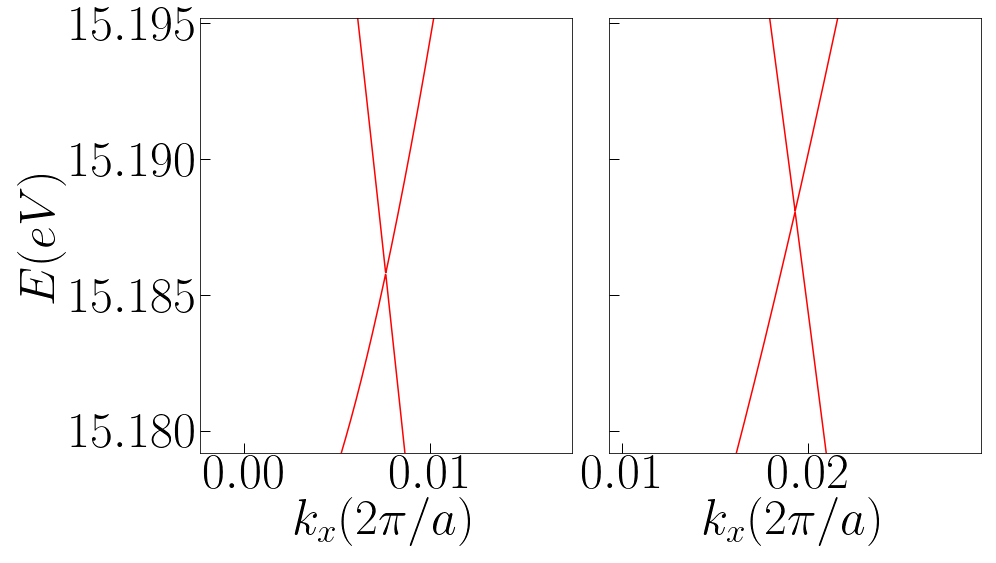}
    \caption{Tilted Weyl cone bandstructures near (a) a W1 node at (0.0076, 0.5140, 0)  and (b) a W2 node at (0.0193, 0.2818, 0.5899) in units of ($2\pi/a, 2\pi/a, 2\pi/c$ ) in the Brillouin zone of TaAs}
    \label{fig:tilt}
\end{figure}

In this section, we will  apply the previous formulation to   find the condition for the LL collapse [Eq.~\eqref{eq:crit_principal}] in TaAs, the best-known Weyl semimetal~\cite{lv2015experimental}. The TaAs system has a body-centered tetragonal structure with lattice parameters $a = 3.437$~\AA\ and $c = 11.656$~\AA\ [Fig. \ref{fig:structure} (a)]. In the Brillouin zone of TaAs, there are 24 Weyl nodes in total, among which 8 nodes are located in $k_z=2\pi/c$ plane (W1 nodes) and the other 16 nodes are located outside this plane (W2 nodes) [Fig. \ref{fig:structure} (b)]. Figure~\ref{fig:tilt} shows the tilted Weyl cones in the bandstructure of TaAs.

\begin{table*}[]
    \centering
    \begin{tabular}{c||c c c |c c c |c c c |c}
         \hline
            Weyl node& $w_x$& $w_y$& $w_z$&$v_{x'}$ &$v_{y'}$&$v_{z'}$ & $\hat{x}'_{\rm p}$ & $\hat{y}'_{\rm p}$& $\hat{z}'_{\rm p}$ \\
            \hline
            W1(A) &-1.194 & -0.855 & 0 & 4.053 & 2.145 & 0.238 &(0.977, 0.214, 0)&(-0.214, 0.977, 0)&(0, 0, 1) \\
            W2(A$'$) & -0.892 & 0.850 & 1.364 &4.671 & 1.087 &2.079 & (-0.475, -0.707, 0.524) &(-0.651, -0.119, -0.750) &(-0.593, 0.679, 0.404) \\
            \hline
    \end{tabular}
    \caption{Bandstructure parameters of TaAs Weyl cones: the tilt velocities in the lab frame, anisotropic Fermi velocities in the principal frame, and the corresponding principal axes. The tilt velocities and anisotropic Fermi velocities are in units of  $10^5$ m/s. We consider the W1 node at (0.0076, 0.5140, 0)  and W2 node at (0.0193, 0.2818, 0.5899) in units of ($2\pi/a, 2\pi/a, 2\pi/c$ ) in the Brillouin zone of TaAs.
    }
    \label{tab:weyl}
\end{table*}

Table~\ref{tab:weyl} shows $\textbf{w}$, the tilt velocity [Eq.~\eqref{eq:tilt}], $v_x'$, $v_y'$, and $v_z'$, the principal values of $\textbf{U}$ [Eqs.~\eqref{eq:v_OU}-\eqref{eq:D}], and $\hat{x}'_{\rm p}$, $\hat{y}'_{\rm p}$, and $\hat{z}'_{\rm p}$, the corresponding principal axes, obtained by following the procedure detailed in Sec.~\ref{sec:anisotropy}. The values for $\textbf{w}$ and $v_x'$, $v_y'$, and $v_z'$ are in good agreement with those reported in a previous study~\cite{grassano2020influence}.

\begin{table}[]
    \centering
    \begin{tabular}{c||c c c |c|c}
         \hline
            Weyl node& $b_{xzz}$& $b_{yzz}$& $b_{zzz}$&$g$
            \\
            \hline
            W1\,(A) &-0.136 & 2.125& 0 & -0.327 
            \\
            W2\,(A$'$)  &0.053& -0.097& 1.196 &0.690 
            \\
            \hline
    \end{tabular}
    \caption{Weyl node shift vectors in the lab frame and the energy shift per unit strain for tensile strain along $z$: the W1 and W2 nodes are, respectively, at (0.0076, 0.5140, 0) and (0.0193, 0.2818, 0.5899), in units of ($2\pi/a, 2\pi/a, 2\pi/c$) in the Brillouin zone of TaAs. The Weyl node shift vector is shown in units of $\textrm{\AA}^{-1}$, and the energy shift per unit strain in units of eV. Note that W1~(W2) node in this table means B~(B$'$) node in Fig.~\ref{fig:criterion_zstrain}.
    }
    \label{tab:strain1}
\end{table}

\begin{table}[]
    \centering
    \begin{tabular}{c||c c c |c }
         \hline
            Weyl node& $b_{xxx}$& $b_{yxx}$& $b_{zxx}$&$g$
            \\
            \hline
            W1\,(A) &-0.797& -0.047& 0& 0.520 
            \\
            W1$'$\,(B) &0.236& -0.175& 0 & 1.374
            \\
            W2\,(A$'$) &1.442& -0.143& 0.605&-2.682 
            \\
            W2$'$\,(B$'$)  &0.252&-0.607&-1.415&-2.628 
            \\
            \hline
    \end{tabular}
    \caption{Weyl node shift vectors in the lab frame and the energy shift per unit strain for tensile strain along $x$: the W1, W1$'$, W2, and W2$'$ nodes are, respectively, at (0.0076, 0.5140, 0), (0.5140, 0.0076, 0), (0.0193, 0.2818, 0.5899), and (0.2818, 0.0193, 0.5899) in units of ($2\pi/a, 2\pi/a, 2\pi/c$) in the Brillouin zone of TaAs. The Weyl node shift vector is shown in units of $\textrm{\AA}^{-1}$), and the energy shift per unit strain in units of eV. Note that W1~(W2) node in this table means B~(B$'$) node in Fig.~\ref{fig:criterion_xstrain}, and W1$'$~(W2$'$) means A~(A$'$) therein.
    }
    \label{tab:strain2}
\end{table}

\begin{figure*}
    \includegraphics[width=0.8\textwidth]{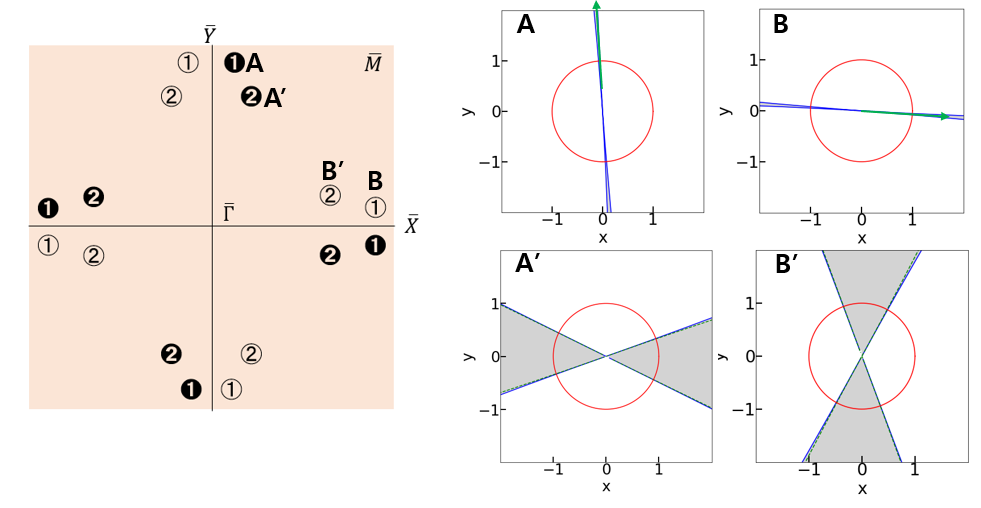}
    \caption{Each polar plot shows the left hand side of Eq.~\eqref{eq:crit_principal} for the corresponding Weyl node (specified in the left panel) as its radius as a function of the direction of the strain gradient vector ${\bf a}$ [Eq.~\eqref{eq:u_xx}] in the blue or green curve; for a positive value, the curve is solid blue, and for a negative value, the curve is dashed green. The radius of red circles is~$1$. The green arrows in panels A and B show the direction of the strain-gradient vector $\bf{b}$ at each node. Here, the strain is applied along $z$. The direction along which the blue curve is either inside the red circle or is dashed green is where the LL collapse occurs. The region where the collapse occurs is painted in gray. }
    \label{fig:criterion_zstrain}
\end{figure*}

Tables~\ref{tab:strain1} and~\ref{tab:strain2} show the calculated values for the strain gradient vector, $\textbf{a}$, and the energy shift per unit strain, $g$. By using these parameters, together with $\textbf{D}$ and $\tilde{\textbf{R}}$ matrices shown in Tab.~\ref{tab:weyl}, we can calculate the renormalized vectors $\textbf{w}'$, $\textbf{a}'$, and $\textbf{b}'$ in Eq.~\eqref{eq:primed_vectors}, and using these vectors, finally obtain the criterion on the direction of the strain gradient vector, $\textbf{a}$, for the collapse of LLs [Eq.~\eqref{eq:crit_principal}].

\begin{figure*}
    \includegraphics[width=0.85\textwidth]{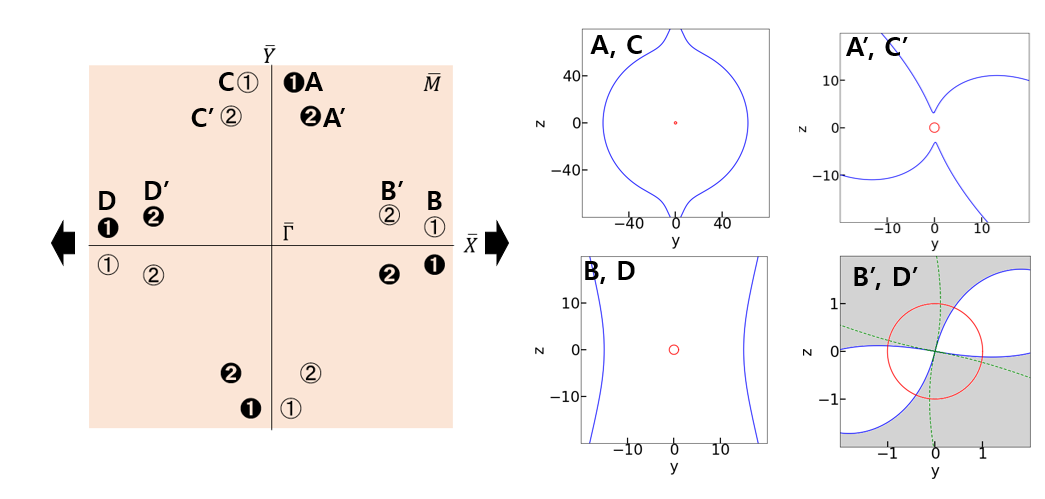}
    \caption{Each polar plot shows the left-hand side of Eq.~\eqref{eq:crit_principal} for the corresponding Weyl node (specified in the upper left panel) as its radius as a function of the direction of the strain gradient vector ${\bf a}$ [Eq.~\eqref{eq:u_xx}] in the blue or green curve; for a positive value, the curve is solid blue, and for a negative value, the curve is dashed green (the lower-right panel for B$'$ and D$'$). The radius of red circles is~$1$. Here, the strain is applied along $x$. The direction along which the blue curve is either inside the red circle or is dashed green is where the LL collapse occurs. The region where the collapse occurs is painted in gray. }
    \label{fig:criterion_xstrain}
\end{figure*}

Figures~\ref{fig:criterion_zstrain} and~\ref{fig:criterion_xstrain} show the left-hand side of the criterion in Eq.~\eqref{eq:crit_principal} at a given strain gradient vector $\mathbf{a}$ [Eq.~\eqref{eq:u_xx}], for strains along $z$ and $x$, respectively. If this value is lower than~1, the LLs arising from a specific Weyl node collapse.  To concentrate on bending by an external stress, we confine the direction of $\textbf{a}$ to be orthogonal to the direction of the strain. First, the condition on $\mathbf{a}$ for the collapse of LLs depends on the Weyl node and the strain tensor (here, along $x$ or along $z$). For example, the collapse cannot occur at Weyl nodes A-D, A$'$, and C$'$ when the tensile strain is applied along $x$  [Fig.~\ref{fig:criterion_xstrain}]. On the other hand, In the case of A$’$, B$’$ in Fig.~\ref{fig:criterion_zstrain}  and B$’$, D$’$ in Fig.~\ref{fig:criterion_xstrain}, we can easily check the collapse.

Moreover, the criteria for different Weyl nodes connected by a symmetry operation of TaAs are connected by the same symmetry operation.
For example, node A (A$'$) is connected to node B (B$'$) by a C$_4$ rotation in momentum space (thanks to a screw rotation $4_1$ in real space involving a quarter-lattice-parameter translation along $z$) and a mirror reflection with respect to the $k_y-k_z$ plane (thanks to the mirror reflection with respect to the $yz$ plane in real space). Thus, for strain along $z$, the condition for LL collapse for node A (A$'$) is connected to that for node B (B$'$) by a mirror reflection with respect to the $y=x$ plane (Fig.~\ref{fig:criterion_zstrain}). Also, nodes A, A$'$, B, and B$'$ are connected to nodes C, C$'$, D, and D$'$, respectively, by a mirror reflection with respect to the $k_y-k_z$ plane (thanks to the mirror reflection with respect to the $yz$ plane). Therefore, the two nodes in each pair have the {\it same} condition for the collapse of LLs for strain along $x$ (Fig.~\ref{fig:criterion_xstrain}; note that in this case the strain gradient vector {\bf a} is confined within the $yz$ plane); however, the condition for node A (A$'$) is not connected to that for node B (B$'$) due to the strain-induced breaking of the $4_1$ symmetry. Notably, the collapse always occurs when the vector $\textbf{a}'$ and $\textbf{b}'$ are parallel to each other because $\textrm{sin}(\theta_{\textbf{a}',\textbf{b}'})=0$ in Eq.~\eqref{eq:crit_principal}. This condition is equivalent to $\mathbf{a}$ and $\mathbf{b}$ being parallel to each other [see Eq.~\eqref{eq:H_complex}]; we can check this reasoning in the panels for nodes A and B in Fig.~\ref{fig:criterion_zstrain}. 

Although the criterion for the pseudo-Landau level collapse does not depend on the magnitude of strain, the formation of the pLLs does. In the Landau quantization experiments, we need a sufficiently high magnetic field to resolve LLs because of the thermal fluctuation and impurities. Likewise, pLLs will be observable only if the pseudo-magnetic field is sufficiently high. The required pseudo-magnetic field depends on the quality of the sample. Landau levels in graphene and NbAs are resolved when the magnetic field is higher than $3 \sim 10$ T~\cite{li2007observation,yuan2018chiral}, respectively. Likewise, we think the pseudo-magnetic field should be larger than $3 \sim 10$ T in our case.

We can present the roughly estimated value of required strain gradient to make $3 \sim 10$ T of pseudo-magnetic field. When we set the scale of $b_{ijk}$ is of the order of 1/$\textrm{\AA}$ (see Table.~\ref{tab:strain1} and ~\ref{tab:strain2}), the scale of pseudo magnetic field is $\tilde{\textbf{B}}= \textbf{a} \times \textbf{b} \approx ab \approx a \times 1/\textrm{\AA}$  (See Eq.~\eqref{bfield2}), and it should have the same scale as $e/\hbar \times 3 \sim 10$ (T). Finally, the strain gradient should be higher than $0.003 \sim 0.01 \%$ / Å to make a pseudo magnetic field of $3 \sim 10$ T. In this range of strain gradient, the collapse of pLLs could be observed if conditions shown in Fig.~\ref{fig:criterion_zstrain} and Fig.~\ref{fig:criterion_xstrain} are satisfied.

Furthermore, there are some experimental methods which can make strain gradient to observe the collapse of pLLs. The first method uses a piezoelectric device~\cite{cenker2022reversible}. By this piezoelectric device, large (over 1\%) strain and strain gradient could be generated, and a strain gradient could be generated, too~\cite{StrainLL19}. Also, a strain gradient can be induced during the cleaving of a layered material, and it can be checked by the ripples in scanning tunneling microscopy images. An experimental study has shown the generation of a pseudo magnetic field equivalent to 3 T in Re-doped $\rm{MoTe}_2$, a Weyl semimetal. Another method to generate a strain gradient is to bend the nanoribbons~\cite{zheng2021strain}. Nanoribbons could be bent under an optical microscope using a glass tip, and the bending shape could be consolidated by an atomic layer deposition system. By this method, 0.002 \% / Å of strain gradient can be made, which is the same scale as the value to make a pseudo magnetic field of $3 \sim 10$ T.

\section{CONCLUSIONS AND OPEN QUESTIONS} \label{sec:conclusions}

Weyl semimetals have Dirac cones in their electronic band structure, which allow fascinating relativistic phenomena to be realized in table-top experiments. Due to the relativistic nature of the WSMs, Landau bands formed by an external magnetic field are different than those of a standard electron gas and can collapse when perpendicular electric and  magnetic fields are applied. Also due to the Dirac nature of the quasiparticles, strain of the ion lattice couples to the electronic density in the form of vector and scalar potentials and pseudo-electromagnetic fields can be induced by strain. Pseudo-LLs have already been observed~\cite{kamboj2019generation} and additional strain can lead to the collapse of these elastic LLs.

In this work, we have investigated the electronic structure and the condition for the collapse of LLs in realistic Weyl semimetals taking into account the full electronic structure. We have extended previous results on the LL collapse done on minimal models of WSMs to the more realistic case of anisotropic and tilted Weyl cones. We have also developed a formalism to treat the lattice strain in these realistic situations. 

Finally, using our theory and first-principles calculations we derived the criterion for the strain-induced collapse of LLs in TaAs, a prototypical Weyl semimetal. As discussed in Ref.~\cite{arjona2017collapse}, the criterion is determined by material parameters: the anisotropic Fermi velocities, tilt velocities, and the energy shift and Weyl node shift vector per unit strain. We found that  the criterion for the collapse of the LLs depends on the direction of the strain gradient vector, but not on its magnitude. 
However, in order for the collapse of LLs to be observed, the pseudo-LLs should be discernible, which determines the lower bound for the magnitude of the strain.

The results of this work can easily be applied to the cases of other materials and set a solid basis for the experimental observation of this novel effect. 

A very interesting possibility is to study the interplay between real and pseudo-electromagnetic fields. In particular, it can be seen if real LLs induced by an external magnetic field in a given direction can collapse by the introduction of a pseudoelectric field in the perpendicular direction. 

\subsection{COMPUTATIONAL DETAIL} \label{sec:calculation}

We calculated the electronic structure of pristine and strained TaAs using density functional theory as implemented in the Quantum-ESPRESSO package~\cite{giannozzi2009quantum}. From these results, we calculated the required material parameters for our theory. We used fully-relativistic, norm-conserving pseudopotentials~\cite{hamann2013optimized, schlipf2015optimization} to treat spin-orbit coupling, and approximated the exchange-correlation energy by the scheme of Perdew, Burke, and Ernzerhof~\cite{perdew1996generalized}. The Brillouin zone was sampled with a a 10 $\times$ 10 $\times$ 10 Monkhost-Pack ~\cite{monkhorst1976special} k-point mesh, and the kinetic energy cutoff was set to 100 Ry. Fig. \ref{fig:structure} (c) shows the calculated band structure of TaAs. We interpolated the electronic bandstructure using Wannier90~\cite{mostofi2008wannier90, mostofi2014updated} and used Wanniertools~\cite{wu2018wanniertools} to find Weyl nodes.

\begin{acknowledgments}
We thank Ji Hoon Ryoo and Massimiliano Stengel for fruitful discussions. This work originated during the visits of  C.\,-H.\,P to Centro de Física de Materiales, Universidad del País Vasco, and  M.\,A.\,H.\,V. to the Donostia International Physics Center (DIPC) whose kind support is deeply appreciated.
Y.\,-J.\,L. and C.\,-H.\,P. were supported by the Institute for Basic Science (No. IBSR009-D1) and by the Creative-Pioneering Research Program through Seoul National University and M.\,A.\,H.\,V. is supported by the project PGC2018-099199-B-I00 (MCIU/AEI/FEDER, UE). Computational resources were provided by KISTI Supercomputing Center (Grant No. KSC-2020-INO-0078).
\end{acknowledgments}

\bibliography{main}

\end{document}